\documentclass[12pt]{article}
 \pdfoutput=1
 \usepackage{graphicx}
  \usepackage{hyperref}
  \usepackage{epsfig}
  \usepackage{amsmath}
  \usepackage{amssymb}
  \usepackage{color}
  \usepackage[nosort]{cite}
  \newlength{\abstractwidth}
  \newcommand{\be}{\begin{equation}}
  \newcommand{\ee}{\end{equation}}
  \newcommand{\tr}{\text{tr}}
  \renewcommand{\title}[1]{\vbox{\center\bf{\Large{#1}}}\vspace{5mm}}
  \renewcommand{\author}[1]{\vbox{\center#1}\vspace{5mm}}
  \newcommand{\address}[1]{\vbox{\center\em#1}}
  \newcommand{\email}[1]{\vbox{\center\tt#1}\vspace{5mm}}
  \renewcommand{\k}{\mathbf{k}}
  \newcommand{\p}{\mathbf{p}}
  \newcommand{\q}{\mathbf{q}}

\begin{document}

\begin{titlepage}
\begin{center}
\hfill \\
\hfill \\
\vskip 1cm

\title{Many-body chaos at weak coupling}

\author{Douglas Stanford}

\address{
 School of Natural Sciences, Institute for Advanced Study \\
 Princeton, NJ, USA

}

\email{stanford@ias.edu}

\end{center}
  
  \begin{abstract}
 The strength of chaos in large $N$ quantum systems can be quantified using $\lambda_L$, the rate of growth of certain out-of-time-order four point functions. We calculate $\lambda_L$ to leading order in a weakly coupled matrix $\Phi^4$ theory by numerically diagonalizing a ladder kernel. The computation reduces to an essentially classical problem.
\end{abstract}
   \end{titlepage}

\tableofcontents

\baselineskip=17.63pt

\section{Introduction}
Non-time-ordered four point functions can be used to diagnose chaos in many-body quantum systems  \cite{larkin,Almheiri:2013hfa,SS,Shenker:2013yza,Roberts:2014isa,kitaev,Shenker:2014cwa,Roberts:2014ifa}. For example, we can understand the butterfly effect as the statement that for rather general operators $W,V$, the thermal expectation value of the square of a commutator
\be
c(t) = \langle [W(t),V] [W(t),V]^\dagger\rangle_\beta
\ee
should become large at late time, of order $2\langle V^2\rangle_\beta \langle W^2\rangle_\beta$. The way this function grows can be interesting. At least for simple operators in large $N$ systems, one expects a long period of exponential growth \cite{kitaev},
\be
c(t) \propto \frac{1}{N^2} e^{\lambda_L t}
\ee
plus higher orders in $N^{-1}$ that make the function eventually saturate. The rate defined by the exponential, $\lambda_L$, is a measure of the strength of chaos. It satisfies $\lambda_L \le \frac{2\pi}{\beta}$ \cite{Maldacena:2015waa}.

The purpose of this paper is to evaluate $\lambda_L$ in a weakly coupled large $N$ quantum field theory at finite temperature. Ref.~\cite{Shenker:2014cwa} suggested an approach to the weak coupling calculation, based on analogy to the BFKL \cite{Kuraev:1977fs,Balitsky:1978ic} analysis of high energy scattering in gauge theories. In this paper we set up and carry out this calculation for a simple model system. Specifically, we consider the theory of a single Hermitian matrix field $\Phi_{ab}$ in four spacetime dimensions, with the Lagrange density\footnote{In v1 of this paper, the interaction term was missing the factor of 2. This error was noticed by Tyler Guglielmo and Phuc Nguyen during work on \cite{Fischler:2021rxy} and also by the authors of \cite{Grozdanov:2018atb}.}
\be
\mathcal{L} = \frac{1}{2}\tr\left( \dot{\Phi}^2 -(\nabla\Phi)^2 - m^2 \Phi^2 - 2g^2 \Phi^4\right).
\ee
The `t Hooft coupling is defined as $\lambda = g^2 N$. The goal is to compute $\lambda_L$ to leading order in $\lambda$. We will take a very direct approach, evaluating a subset of thermal Feynman diagrams for an index-averaged and spatially-averaged version of the squared commutator:
\be\label{tocompute}
C(t) = \frac{1}{N^4}\sum_{aba'b'}\int d^3\mathbf{x} \ \tr\Big( \sqrt{\rho}\, [\Phi_{ab}(t,\mathbf{x}),\Phi_{a'b'}]\sqrt{\rho}\, [\Phi_{ab}(t,\mathbf{x}),\Phi_{a'b'}]^\dag\Big).
\ee
Here $\rho = \rho(\beta)$ is the thermal density matrix at inverse temperature $\beta$. Splitting $\rho$ into the two $\sqrt{\rho}$ factors amounts to putting the two commutators on opposite sides of the thermal circle. It is a choice that does not affect $\lambda_L$, but that makes some of the equations below a little simpler.

\begin{figure}[t]
\begin{center}
\includegraphics[scale = .9]{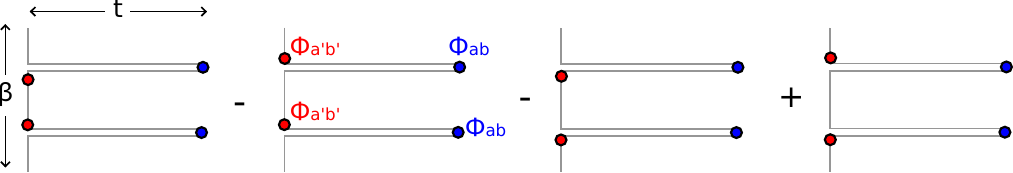}
\caption{The squared commutator (\ref{tocompute}) can be expanded to four terms represented by the path integral contours shown. The vertical segment (ends should be identified) represents the imaginary-time circle, and the horizontal folds implement the real time evolution to produce $\Phi_{ab}(t)$. The two folds are separated by half of the thermal circle.}\label{fourterms}
\end{center}
\end{figure}

To generate the perturbation theory for $C(t)$, it is helpful to think about expanding out the two commutators to give four terms. Each term could be computed by a particular analytic continuation of the Euclidean correlator. Equivalently, we can represent the four terms by path integral contours in complex time, where we append some real-time folds to the Euclidean thermal circle. This is illustrated in Fig.~\ref{fourterms}. In principle, for each term we should follow the usual procedure of expanding down powers of the interaction vertex, integrating them along the contour and connecting fields by contour-ordered propagators. 

The quantity $\lambda_L$ is defined by the asymptotic rate of growth of the $N^{-2}$ term in $C(t)$. In order to compute this, we can restrict to planar diagrams. We can also restrict the region of integration for the interaction vertices to the real-time folds. The integral over the thermal circle implements corrections to the thermal state; such corrections would be important for getting the exact $C(t)$, but we believe that they do not affect the spectrum of growth exponents. These two simplifications would be valid at any value of the coupling, but they are not enough to make the problem tractable. To compute $\lambda_L$ to leading order in the coupling, we can make another simplification, which is to keep only the fastest-growing function of time at each order. Roughly, we will sum all powers of $\lambda^2 t$, but ignore terms proportional to e.g. $\lambda^3 t$ or $\lambda$. This simplification restricts the class of diagrams that we need to consider, and it also allows us to get by with simplified versions of the diagrams that we keep. The structure is very similar to the leading-log approximation in high energy scattering, but with $t$ playing the role of $\log s$ \cite{Shenker:2014cwa}.

The diagrams that must be summed consist of a set of dressed ladder diagrams. The rate of growth of the sum of ladders can be determined by finding the largest eigenvalue of a one-dimensional integral equation, which we diagonalize numerically. In the case where the bare mass $m$ is nonzero but small compared to the temperature, we find
\be
\lambda_L \approx 0.025\frac{\lambda^2}{\beta^2 m} \hspace{20pt} m\beta \ll 1.
\ee
The fact that $\lambda_L$ is proportional to $1/m$ indicates that the important degrees of freedom are the highly populated, frequently colliding low energy quanta with $E\sim m$, not the thermal scale quanta that one might have expected. Naively, this result diverges for a massless field, but if the tree level mass vanishes we must include the one-loop thermal mass $m^2_{th} = 2\lambda/3\beta^2$, giving
\be
\lambda_L \approx 0.031\frac{\lambda^{3/2}}{\beta} \hspace{20pt} m = 0.
\ee
This is still small, but it is parametrically enhanced relative to the naive $\lambda^2$ scaling.

In addition to the parallels to BFKL, our calculation shares much in common with the analysis by Jeon \cite{jeon} of the shear viscosity in weakly coupled $\phi^4$ theory. Both the work of Jeon and the review of BFKL by Forshaw and Ross \cite{Forshaw:1997dc} were very useful guides. 

Kitaev \cite{kitaev2} has computed $\lambda_L$ in a strongly-coupled fermion quantum mechanics similar to the Sachdev-Ye model \cite{ Sachdev:1992fk, parcolletgeorges,Sachdev:2015efa}. The diagrams are structurally similar to the ones that we study in this paper, but in that context they give the exact $O(1/N)$ answer as a function of the coupling, saturating the bound $2\pi/\beta$ as the coupling goes to infinity.

\section{The ladder diagrams}\label{ladder}
In this main section of the paper, we will study a set of ladder diagrams for $C(t)$ and derive an eigenvalue equation that determines the growth rate $\lambda_L$ at leading order in the coupling $\lambda$. This equation arises from the fact that an infinite ladder is not changed if we add one extra rung. There are two slightly subtle points in the analysis. The first is related to the fact that we are doing perturbation theory on a pair of folded time contours, and the interaction vertices should be integrated over both sides of each fold. The sum over the two sides turns the side rails in the ladder diagrams into retarded propagators, while the rungs remain Wightman correlators. The second subtlety is that we have to include self-energy corrections for the retarded propagators. In order to explain both of these points, we will go through the first couple orders of perturbation theory explicitly, in \S\ref{free}, \S\ref{first} and \S\ref{second}. We will then analyze the ladder diagrams in \S\ref{higher}, discussing corrections in \S\ref{correctionsSec}.

\subsection{Free propagators}
For the computations below, we will need two types of correlation functions: the retarded propagator $G_R$ and a Wightman function $\tilde{G}$ with the operators separated by half of the thermal circle. The functions are defined by
\begin{align}
\delta_{ab'}\delta_{ba'}G_R(\mathbf{x},t)&= \theta(t) \tr\Big(\rho\,  [\Phi_{ab}(\mathbf{x},t),\Phi_{a'b'}]\Big)\\
\delta_{ab'}\delta_{ba'}\tilde{G}(\mathbf{x},t)&=\tr\Big(\sqrt{\rho}\, \Phi_{ab}(\mathbf{x},t)\sqrt{\rho}\,\Phi_{a'b'}\Big) .
\end{align}
Here and elsewhere in the paper, a field operator without spacetime coordinates is assumed to be at the origin $\Phi\equiv \Phi(0,0)$.

In momentum space, it is often useful to express these propagators in terms of the spectral function $\rho(\omega,|\k|)$. (We will distinguish the spectral function $\rho(k) = \rho(k^0,|\k|)$ from the thermal density matrix $\rho$ by making the arguments explicit.)
\begin{align}
G_R(k) &= i\int \frac{d\omega}{2\pi}\frac{\rho(\omega, |\k|)}{k^0 - \omega + i\epsilon}\\
\tilde{G}(k) &= \frac{\rho(k)}{2\sinh \frac{\beta k^0}{2}}.
\end{align}
For a free field, we have $\rho(\omega,\k) = \frac{\pi}{E_{\k}}[\delta(\omega-E_{\k}) - \delta(\omega + E_{\k})]$, and therefore
\begin{align}
G_R(k) &= \frac{i}{2E_{\k}}\left(\frac{1}{k^0 - E_{\k}+i\epsilon} - \frac{1}{k^0 + E_{\k} + i\epsilon}\right)\label{retardedG}\\
\tilde{G}(k) &= \sum_{s = \pm}\frac{\pi\delta(k^0-sE_\k)}{2E_\k\sinh\frac{\beta E_\k}{2}}.\label{tildeG}
\end{align}
We will now use these propagators to discuss the first couple of orders of perturbation theory for $C(t)$.

\subsection{Order $\lambda^0$}\label{free}
In the free theory, we simply sum over the three ways of contracting the four operators on the contours of Fig.~\ref{fourterms}. Two of these cancel when we take the sum of the terms to form the square of the commutator. The only nonvanishing contraction is the one in which the two fields on the bottom fold are contracted with each other, and likewise for the top fold. For this pattern of contractions, the sum over the four terms gives
\be
C_{free}(t) = -\frac{1}{N^2}\int d^3\mathbf{x} \, G_R(\mathbf{x},t)^2,
\ee
where the minus sign comes from the $\dagger$ reversing the order of the operators inside the commutator in Eq.~(\ref{tocompute}). 

\subsection{Order $\lambda$}\label{first}
\begin{figure}[t]
\begin{center}
\includegraphics[scale = .9]{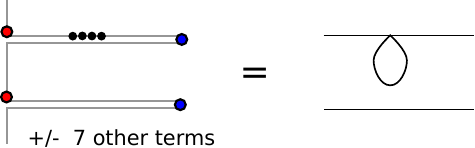}
\caption{When one interaction vertex is integrated over one of the folds, we get a self-energy correction for one of the retarded propagators. The seven other terms on the LHS differ in whether the vertex is on the top piece or bottom piece of the fold, and in the arrangement of the external operators. The sum (with signs) turns all horizontal lines into retarded propagators.}\label{1b}
\end{center}
\end{figure}
The first correction to the free theory comes from a term where we integrate one copy of the interaction vertex over both of the real-time folds. In Fig.~\ref{1b} we show the one-loop self energy diagram that results. Similar diagrams appear at higher orders in perturbation theory as well, decorating all propagators. The effect is a one-loop temperature-dependent correction to the mass of the field. In principle, this can be absorbed by using the ``thermal mass'' $m^2_{th}(\beta)$ in all propagators. However, for nonzero $m$, $\lambda_L$ depends smoothly on the mass, so to leading order in $\lambda$ we can just use the tree-level mass and ignore the one-loop self energy altogether. The exception is the case where the tree-level mass is zero; there we must include the thermal mass. We work out the value in appendix \ref{oneloop}:
\be
m_{th}^2 = \frac{2\lambda}{3\beta^2} \hspace{20pt}(m = 0).
\ee

\begin{figure}[t]
\begin{center}
\includegraphics[scale = .9]{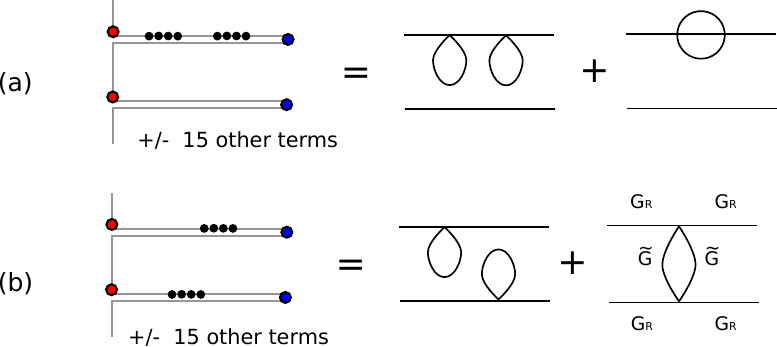}
\caption{When both vertices in the $O(\lambda^2)$ correction are integrated over the same fold, we get the two self-energy diagrams shown in (a). When the vertices are integrated over different folds, we get additional self-energy corrections and then, finally, the one-rung diagram shown at right in (b). For this diagram, we emphasize which propagators are retarded and which are Wightman.}\label{setup3}
\end{center}
\end{figure}

\subsection{Order $\lambda^2$}\label{second}
At order $\lambda^2$, we integrate two interaction vertices over the contour. When these vertices are on the same Lorentzian fold, as in Fig.~\ref{setup3}(a), we get the second term in the geometric series of one-loop self energies, as well as the first term of the two-loop self energy. We can account for the two-loop diagram (as well as similar diagrams dressing propagators at higher orders in perturbation theory) by including a $\lambda^2$ self energy correction in the propagators. The real part of the two-loop self energy is a momentum-dependent shift in the mass that can be ignored relative to the tree-level or one-loop mass discussed above. The imaginary part leads to exponential decay of correlation functions due to the finite lifetime of a single particle state; at finite temperature, a particle can be knocked into a different momentum mode by a collision with a thermal excitation. We will see below that this process will contribute to the leading order $\lambda_L$. We can include the relevant effect by modifying the retarded propagators to
\be
G_R(k) \approx \frac{i}{2E_{\k}}\left(\frac{1}{k^0 - E_{\k}+i\Gamma_{\k}} - \frac{1}{k^0 + E_{\k} + i\Gamma_{\k}}\right).
\ee
The computation of the two-loop width $\Gamma_{\k}$ is standard \cite{Parwani:1991gq,jeon}, and we review it in Appendix \ref{twoloop}.

We will also have a configuration in which one interaction vertex is attached to each of the Lorentzian folds. This leads to a qualitatively new diagram, the ``one rung'' diagram, illustrated in Fig.~\ref{setup3}b.
To analyze this diagram we would like to study the Fourier transform $C(\omega)$. This is slightly delicate, because we anticipate that the result of our computation will be an exponentially growing function $C(t)$, for which the Fourier transform does not exist. Instead, we will actually study a Laplace transform $C(\omega) = \int_0^\infty e^{i\omega t}C(t)$ 
where the integral is over positive $t$ only. We can recover $C(t)$ at positive time by taking $C(t) = \int\frac{d\omega}{2\pi}e^{-i\omega t}C(\omega)$ along a contour that runs above all singularities in the complex $\omega$ plane.

\begin{figure}[t]
\begin{center}
\includegraphics[scale = .9]{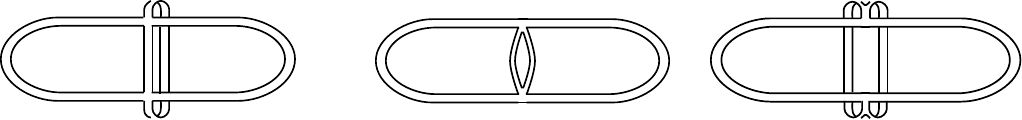}
\caption{The sum over indices in $C(t)$ is equivalent to contracting the external operators in the four point function with semicircle caps. The one rung diagram then has three index structures that each contribute $16N^4$, where $16 = 4\cdot 4$ is a combinatoric factor that arises from the possibility of ``rotating'' each of the vertices in the plane of the diagram. The total factor is $48N^4$; dividing by $N^4$ to turn the sum into an average, we get 48.}\label{setup4}
\end{center}
\end{figure}

Including the combinatoric factor explained in Fig.~\ref{setup4}, the one rung diagram is equal to
\be\label{oiuwreoui}
C_{one \ rung}(\omega) = \frac{1}{N^2}\int \frac{d^4p}{(2\pi)^4}\frac{d^4p'}{(2\pi)^4} G_R(\omega-p)G_R(p)R(p-p')G_R(\omega-p')G_R(p').
\ee
where the rung function $R$ contains the loop integral and the product of Wightman correlators
\be
R(p) = 48\lambda^2\int \frac{d^4\ell}{(2\pi)^4}\tilde{G}(p/2 + \ell)\tilde{G}(p/2 - \ell).
\ee
This function $R$ will be important in what follows. We evaluate the integral explicitly in Appendix \ref{rungapp}. The overall sign in (\ref{oiuwreoui}) is $-i^2 = 1$, where the two factors of $i$ come from the real-time interaction vertex, and the minus sign comes from the $\dagger$ in (\ref{tocompute}).

We would like to understand how to extract the leading large-time behavior of (\ref{oiuwreoui}). It is helpful to begin by understanding the expression with the free propagators. In this case, the leading large-time behavior of $C_{one \ rung}(t)$ is linear in $t$. In frequency space, this comes from a double pole in $\omega$. Let us see how we get this behavior. The above expression contains two pairs of retarded propagators. With the free propagators, the first pair $G_R(p)G_R(\omega-p)$ is
\be\label{twoGR}
-\frac{1}{4E_{\p}^2}\left(\frac{1}{p^0 - E_{\p}+i\epsilon} - \frac{1}{p^0 + E_{\p} + i\epsilon}\right)\left(\frac{1}{\omega - p^0 - E_{\p}+i\epsilon} - \frac{1}{\omega - p^0 + E_{\p} + i\epsilon}\right).
\ee
The integral of the complete expression over $p^0$ will have a contribution from taking the residues of these poles. We get terms proportional to $\omega^{-1}$, to $(\omega + 2E_{\p})^{-1}$ and to $(\omega - 2E_{\p})^{-1}$. The first term is the important one, because when we combine it with a similar term from the second pair of retarded propagators, we will get the desired double pole $\omega^{-2}$. The lesson is that we can get the correct large-time behavior by making the replacement
\be\label{replacement}
G_R(p) G_R(\omega-p) \rightarrow -\frac{\pi i}{2 E_{\p}^2}\frac{\delta(p^0 - E_{\p}) + \delta(p^0 + E_{\p})}{\omega + i\epsilon}.
\ee

Now we add back in the self energy correction, replacing the $i\epsilon$ factors in (\ref{twoGR}) with $i\Gamma_{\p}$. The shift in the location of the poles will affect the on-shell delta functions, but only by a small amount that we can ignore at leading order. The only effect that we need to include is to modify the replacement as
\be\label{replacement2}
G_R(p) G_R(\omega-p) \rightarrow -\frac{\pi i}{2 E_{\p}^2}\frac{\delta(p^0 - E_{\p}) + \delta(p^0 + E_{\p})}{\omega + 2i\Gamma_{\p}}.
\ee
A convenient feature is that the delta functions in this substitution, together with the delta functions in the Wightman functions $\tilde{G}$, force each of the four momenta $p,p'$ and $(p-p')/2\pm \ell$ to be on shell. This will be important in comparing to a classical calculation below. Notice also that we are supposed to sum over positive and negative energy for each of the momenta.

\subsection{Higher orders}\label{higher}
\begin{figure}[t]
\begin{center}
\includegraphics[scale = .9]{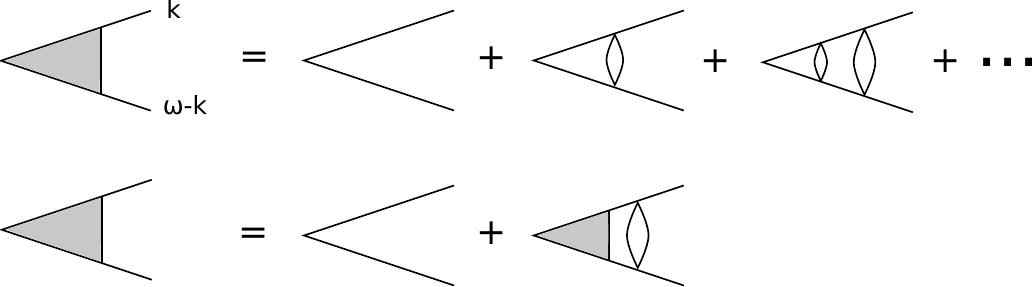}
\caption{Ladder diagrams for $f(\omega,k)$ are shown. In these diagrams, diagonal lines are dressed retarded propagators, and the loops are Wightman correlators $\tilde{G}$. Frequency $\omega$ flows into the diagram from the left corner, which has an implied sum over momenta. The shaded blob represents the full $f$. It satisfies the recursion relation shown on the bottom line. At large $t$, the zero-rung term on the right hand side can be ignored, so we have a homogeneous equation that states that the large-time behavior of the sum is unchanged if we add one extra rung.}\label{setup5}
\end{center}
\end{figure}

At higher orders in $\lambda$ the leading diagrams consist of dressed ladders. To sum these diagrams, it is useful to define the function $f$ through
\be
C(\omega) = \frac{1}{N^2}\int \frac{d^4p}{(2\pi)^4}f(\omega,p).
\ee
The sum of ladders illustrated in Fig.~\ref{setup5}.
satisfies a simple integral equation
\be\label{integrale}
f(\omega,p) = -G_R(p)G_R(\omega-p)\left[1 +\int\frac{d^4k}{(2\pi)^4} R(k-p)  f(\omega,k)\right].
\ee
This is an inhomogeneous equation. At large time we expect $f(t,p)$ to be growing, whereas the homogeneous term will be decaying. We can therefore get the correct large-time rate of growth by dropping the homogeneous term. We also substitute in the approximate form for $G_R$ discussed above. The integral equation then becomes
\be\label{homog}
(-i\omega +2\Gamma_\p)f(\omega,p)\approx \frac{\pi}{E_\p}\delta(p_0^{2}- E_\p^2)\int\frac{d^4k}{(2\pi)^4} R(k-p)  f(\omega,k).
\ee
Because of the on-shell delta functions in all pairs of retarded propagators, $f(\omega,p)$ will be entirely supported for on-shell $p$. We therefore write
\be
f(\omega,p) = f(\omega,\p)\delta(p_0^{2} - E_\p^2),
\ee
where $f(\omega,p)$ and $f(\omega,\p)$ are distinguished by their second argument. Substituting into (\ref{homog}), we get
\be\label{integralequation}
-i\omega f(\omega,\p) = -2\Gamma_\p f(\omega,\p) +  \int \frac{d^3k}{(2\pi)^3}\,m(\k,\p)f(\omega,\k)
\ee
where the kernel $m(\k,\p)$ is given in terms of the rung function $R$ as
\be
m(\k,\p)= \frac{R(k_+) + R(k_-)}{4E_\k E_\p}\hspace{20pt} k_\pm = (E_\k\pm E_\p,\k-\p).
\ee
For on-shell $k $ and $ p$, one can show that $R(k_+)$ comes only from the first term in equations (\ref{twoterms}) and (\ref{runganswer}) while $R(k_-)$ comes only from the second term.

As shown in (\ref{jklasd}), we can actually write the decay rate $\Gamma_\p$ in terms of the rung matrix, so the whole equation is
\be\label{wholeequation}
-i\omega f(\omega,\p) = \int \frac{d^3k}{(2\pi)^3}\,m(\k,\p)\left(f(\omega,\k) - \frac{\sinh \frac{\beta E_\p}{2}}{3\sinh\frac{\beta E_\k}{2}}f(\omega,\p)\right).
\ee
In real time, this equation is $\frac{d}{dt} f = M f$, where $M$ represents the integral operator on the RHS. The eigenvalues of $M$ determine the spectrum of growth exponents; the largest positive exponent is $\lambda_L$. We do not know how to solve this integral equation analytically, but we can solve it numerically. Appendix \ref{numericsapp} and Fig.~\ref{eigenvectorPlot} give some details on this. The results we find were already reported in the Introduction: for small $m\beta$ we find $\lambda_L \approx 0.025 \lambda^2/m\beta^2$. For larger values of $m\beta$ we find that $\lambda_L$ is exponentially small in $\beta m$.

\subsection{Corrections}\label{correctionsSec}
\begin{figure}[t]
\begin{center}
\includegraphics[scale = .9]{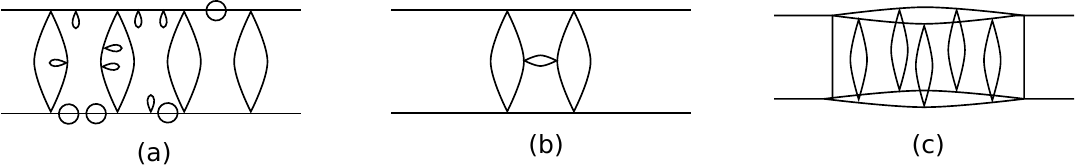}
\caption{(a) shows an example of a ladder diagram dressed by self energy corrections, which is included in our analysis. (b) is an example of a diagram that we expect to correct $\lambda_L$ at order $\lambda^4$. (c) is an example of a double ladder exchange, which we expect to sum to $N^{-4} e^{2\lambda_L t}$.}\label{corrections}
\end{center}
\end{figure}

There are two qualitatively different types of corrections to the analysis above. The first class consists of corrections suppressed by the 't Hooft coupling $\lambda$. These come from other planar diagrams that have more than two powers of $\lambda$ for each power of $t$. An example of such a correction is shown in Fig.~\ref{corrections}(b). By studying decorated ladder diagrams of this type, we expect that one could compute successive corrections to $\lambda_L$ in powers of $\lambda$.

The second class is $1/N$ corrections. These may also correct $\lambda_L$, but they have a second qualitatively different effect, which is to give $N^{-4} e^{2\lambda_L t}$ corrections to $C(t)$. In fact, we expect a full power series in $N^{-2} e^{\lambda_L t}$ that will sum to a function that saturates for large $t$. An example of a diagram that will give this type of correction is shown in Fig.~\ref{corrections}(c). The two ladder sub-diagrams will give a contribution $\propto e^{2\lambda_L t}$, and the overall factor is $1/N^4$. These corrections are analogous to the multi-Pomeron exchange diagrams in high energy scattering, or multi-graviton exchanges for holographic theories.

Actually, the $N$ counting here is a little subtle; we haven't shown the double line diagram, but there is a special subset of index contractions that give a $1/N^2$ piece in the double-ladder diagram. These correspond to taking just one of the three index structures in Fig.~\ref{setup4} for each rung in the ladders. The expected contribution from this subset of contractions would be proportional to $N^{-2}e^{2y t}$, where $y$ is the leading eigenvector of an equation where we divide the first term in (\ref{wholeequation}) by three. We believe that $y = 0$, both based on numerics and intuition to be described in point 2 of the Discussion; this contribution is then $N^{-2}$ uniform in time, and therefore ignorable relative to the $N^{-2}e^{\lambda_L t}$ piece that we calculated.

\section{Discussion}
{\bf 1.} One benefit of going through the weak coupling exercise in this paper is to see that $\lambda_L$ approaches a classical quantity, determined by collision integrals over on-shell momentum. This quantity is related to the following problem: we add a particle of type $i$ to an ensemble of classical particles at time zero, and we ask what the probability is that a particle of type $j$ will be affected some time $t$ later. One can make an analogy to an epidemic \cite{Sekino:2008he,contagion,husekim} spread by the collision of particles. In this picture, the particle we introduce is patient zero, and $\lambda_L$ is the rate of exponential growth of the number of infected particles in the early (pre-saturation) stages of the epidemic.

We can make this more precise by using (\ref{twoterms}) to write (\ref{integralequation}) as
\begin{align}
\frac{d}{dt}f_+(\p,t) = &-2\Gamma_\p f_+(\p,t) \notag\\ &+48\lambda^2\int \frac{d^3k d^3q}{(2\pi)^6}\frac{2\pi\delta(E_\k + E_\p - E_\q - E_{\p - \k - \q})}{2^4E_\k E_\p E_\q E_{\p - \k - \q}(e^{\beta E_{\p - \k - \q}}-1)(e^{\beta E_\q}-1)}f_-(\k,t)\notag\\
&+ 96\lambda^2\int \frac{d^3k d^3q}{(2\pi)^6}\frac{2\pi\delta(E_\k - E_\p + E_\q - E_{\p - \k - \q})}{2^4E_\k E_\p E_\q E_{\p - \k - \q}(e^{\beta E_{\p - \k - \q}}-1)(1-e^{-\beta E_\q})}f_+(\k,t)\label{evolution}
\end{align}
where we defined
\be
f_\pm(\omega,\k) = e^{\mp \beta E_\k/2}f(\omega,\k). 
\ee
We would like to interpret $f_+(\p,t)$ as the expected number of infected particles of type $\p$ at time $t$ and $f_-(\p,t)$ as the expected number of infected holes. This includes the sum over all indices, so the expected number of a given index type would be $f_\pm / N^2$.

The first line in the evolution equation (\ref{evolution}) represents the loss of infected particles of type $\p$ due to collisions with other particles. The third line represents the fact that if an infected particle of type $\k$ collides with a thermal particle, we gain two new infected particles. The factor $96\lambda^2$ is $2|\mathcal{T}|^2$ for the theory we are considering. In such a collision, we also gain one infected hole, for the state $\k$ that is no longer occupied. Similarly, the scattering evolution of an infected hole results in two infected holes and one infected particle. The second line represents this process. The only surprise in the collision integrals of (\ref{evolution}) is the following:  we might have expected to have final state bose factors $1 + n(E_\p) = (1 - e^{-\beta E_\p})^{-1}$ for the new infected particle in the collision integrals of equation (\ref{evolution}). We do not have such factors.

{\bf 2.} From this perspective, we can understand the relative factor of three between the two terms in (\ref{wholeequation}): each collision involving an infected particle results in the loss of one and the creation of three (two particles and one hole). It is because of this factor of three that $\lambda_L$ is positive and $C(t)$ grows.

{\bf 3.} A surprising feature of our result for $\lambda_L$ is the $1/m$ behavior, a type of IR divergence. We can understand where this comes from by looking at (\ref{evolution}), and putting in the massless form $E_\k = |\k|$. If we go to very low momentum we can expand the exponentials in the denominator. We then have six powers of momentum in the denominator and five in the numerator. So if we scale a candidate eigenvector $f(\p)$ toward smaller momentum, we increase the eigenvalue.

This IR effect also shows up in a way in the imaginary part of the self energy. For a massless field, the two-loop decay rate $\Gamma_\p$ is proportional to $1/|\p|$ at small $\p$ \cite{Parwani:1991gq,jeon}. This means that low-momentum modes are experiencing lots of collisions. This does not lead to IR sensitivity for the shear viscosity, because these modes don't carry much momentum, but it does affect $\lambda_L$.

{\bf 4.} The rate of growth of the epidemic is a measure of the many-body butterfly effect, but it is not exactly a Lyapunov exponent for the gas of particles. That would be defined as the growth of an infinitesimal perturbation. In the case at hand, the perturbation is only small in the sense that it initially affects a small fraction of the total number of degrees of freedom $N^2$. 

It would be nice to know if there is some other effective classical description where $\lambda_L$ approaches a Lyapunov exponent. Presumably the holographic dual has this property. Another possibility, in case of very small mass where the relevant modes are highly occupied, would be to study the nonlinear classical field equations in the spirit of recent work on one dimensional matrix systems of \cite{Gur-Ari:2015rcq}. Previous work \cite{Mueller:2002gd,Jeon:2004dh,Mathieu:2014aba} relating the quantum Boltzmann equation (at high occupation number) to nonlinear classical field equations may be useful.

{\bf 5.} As pointed out in \cite{Shenker:2014cwa}, there is another weakly coupled field theory where $\lambda_L$ was computed long ago. This is four dimensional gauge theory on hyperbolic space at temperature $\beta = 2\pi R$. Up to beta functions, which are higher order in the coupling, this problem is equivalent to a high energy scattering problem in the Minkowski vacuum. The BFKL analysis of this problem determines $\lambda_L$ as
\be
\lambda_L = \frac{2\pi}{\beta}(j_0 - 1) = \frac{2\log 2}{\pi \beta}\lambda + O(\lambda^2).
\ee
This is proportional to $\lambda$ at weak coupling, rather than $\lambda^2$ or $\lambda^{3/2}$ as in this paper. The key difference is that the BFKL ladders are cubic ladders because of the cubic Yang-Mills coupling. The vertices come with $\sqrt{\lambda}$, so one rung costs a single factor of $\lambda$. One might wonder whether weakly coupled gauge theory at finite temperature in flat space would have $\lambda_L \propto \lambda$ for the same reason. We believe that it would not, since in flat space one cannot make on-shell ladders out of cubic vertices.

{\bf 6. } It would be nice to compute $\lambda_L$ in other weakly coupled theories, and to explore the behavior of $C(t,\mathbf{x})$ where we do not integrate over the spatial separation of the operators. This might give some insight into the diffusive vs. ballistic behavior in the growth of operators discussed in \cite{Roberts:2014isa,Shenker:2014cwa}.

\section*{Acknowledgements}
I am grateful to Steve Shenker for collaboration during the initial phase of this project, to  Raghu Mahajan, Juan Maldacena, and Vladimir Rosenhaus for discussions, and to Dan Roberts for comments on the draft. This work was supported by the Simons Foundation.

\appendix

\section{Self energies}\label{selfenergy}
We begin with the Euclidean self energy $\Pi(i\omega_n)$, defined at Matsubara frequencies $i\omega_n = 2\pi in/\beta$. We define $\Pi$ so that when we sum the geometric series of self energy insertions, we get the Euclidean correlator
\be
G(i\omega_n,\k)^{-1} =G_0(i\omega_n,\k)^{-1} + \Pi(i\omega_n,\k). 
\ee
The retarded propagator is given by $G_R(k) = -iG(k^0+i\epsilon,\k)$. Plugging this in and using the free expression (\ref{retardedG}), we have the dressed retarded propagator
\be
G_R(k)^{-1} = -i(k^0+i\epsilon)^2 + iE_{\k}^2+i\Pi(k^0+i\epsilon,\k).
\ee
The imaginary part of $\Pi$ is an odd function of $k^0$. If $\Pi$ is small, then we can approximate this form near the on-shell poles as
\be
G_R(k)^{-1} \approx -i(k^0 + i\Gamma_\k)^2 + i(E_\k^2+\Sigma_\k).
\ee
where we have defined
\be
2E_\k \Gamma_\k = -Im[\Pi(E_\k +i\epsilon,\k)]\hspace{20pt} \Sigma_\k = Re[\Pi(E_\k,\k)].
\ee

\subsection{The one-loop self energy}\label{oneloop}
The effect of the one-loop self energy is a shift in the mass. In the case where the field has a nonzero bare mass, this small shift does not affect the leading order calculations in this paper. However, in the case where the field is massless, we have to include the shift. For a massless field, the one-loop self energy is
\begin{align}
\Pi(i\omega_n,\k) &= \frac{8\lambda}{\beta}\sum_n\int \frac{d^3\p}{(2\pi)^3}\frac{1}{\omega_n^2 + \p^2}\\
&=4\lambda \int \frac{d^3\p}{(2\pi)^3}\frac{1}{|\p| \tanh \frac{|\p|\beta}{2}}.
\end{align}
Subtracting off the zero temperature integrand and then doing the integral, we find the thermal mass
\be
m_{th}^2 = \frac{2\lambda}{3\beta^2}.
\ee

\subsection{The imaginary part of the two-loop self energy}\label{twoloop}
To lighten the notation, we will suppress the spatial momenta, restoring them below. Then the two-loop self energy is
\be
\Pi(i\omega_n) = -16\lambda^2 \int_0^\beta d\tau e^{i\omega_n\tau}G(\tau)^3.
\ee
We can use the spectral representation to write the correlators as 
\be
G(\tau) = \int \frac{dk^0}{2\pi}\frac{\rho(k^0)e^{-k^0\tau}}{1-e^{-\beta k^0}}.
\ee
The integral over $\tau$ is now simple. Using $e^{i\omega_n\beta} = 1$, we get
\be
\Pi(i\omega_n) = 16\lambda^2\int \prod_{j = 1}^3\left[\frac{dk_j^0}{2\pi}\frac{\rho(k_j^0)}{1-e^{-\beta k_j^0}}\right]\frac{1-e^{-(k^0_1+k^0_2+k^0_3)\beta}}{i\omega_n - (k_1^0+k_2^0+k_3^0)}.
\ee
We now do the following: {\it (i)} we continue $i\omega_n \rightarrow p^0+i\epsilon$, {\it (ii)} we use $Im[\frac{1}{x+i\epsilon}] = -\pi\delta(x)$ to find the imaginary part, {\it (iii)} we restore the integral over spatial momenta and {\it (iv)} we use the energy conservation delta function to replace the $(1 + e^{-\beta E})$ factors by $2\sinh\frac{\beta E}{2}$. The result is 
\be
-Im[\Pi(p^0+i\epsilon,\p)] = 16\lambda^2 \sinh\frac{\beta p^0}{2}\int \prod_{j = 1}^3\left[\frac{d^4k_j}{(2\pi)^4}\frac{\rho(k_j)}{2\sinh\frac{\beta k_j^0}{2}}\right](2\pi)^4\delta^4(p-k_1-k_2-k_3).
\ee
This can be reduced further, to a one-dimensional integral \cite{Parwani:1991gq,jeon}. However, we are interested in deriving an expression for $\Gamma_\p = -Im[\Pi(E_\p+i\epsilon,\p)]/2E_\p$ in terms of the rung function. To get this, we substitute in for $\rho$ in terms of $\tilde{G}$:
\be
\Gamma_\p = \frac{\sinh\frac{\beta E_\p}{2}}{6E_\p}\int\frac{d^4k}{(2\pi)^4}R(p-k)\tilde{G}(k),
\ee
where the $p$ in the argument of $R$ can be either of $(\pm E_\p,\p)$. We can use the free expression for $\tilde{G}$ (\ref{tildeG}) to write this in a form useful in the main text:
\be\label{jklasd}
\Gamma_\p = \frac{1}{6}\int \frac{d^3k}{(2\pi)^3}\,\frac{\sinh \frac{\beta E_\p}{2}}{\sinh\frac{\beta E_\k}{2}}\,\frac{R(k_+) + R(k_-)}{4E_\p E_\k}\hspace{20pt} k_\pm = (E_\k\pm E_\p,\k-\p).
\ee

\begin{figure}[h]
\begin{center}
\includegraphics[scale = .8]{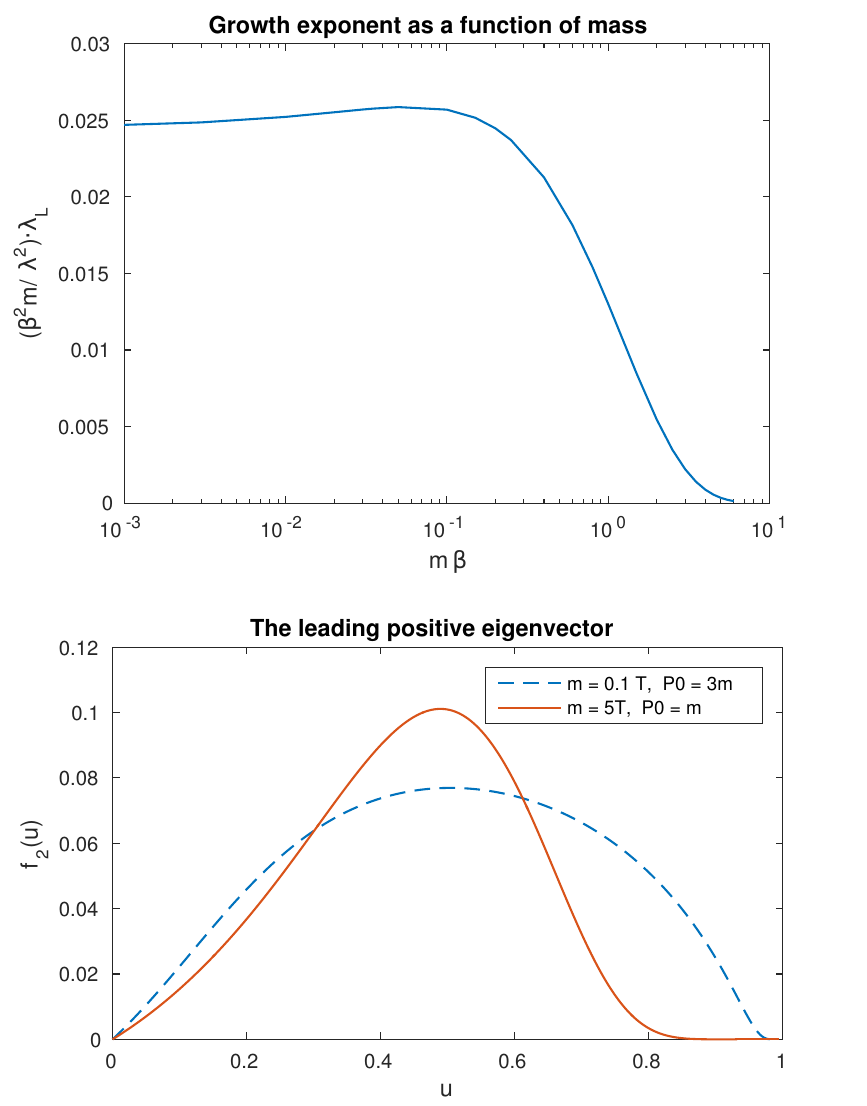}
\caption{{\bf Top:} the combination $(\beta^2m/\lambda^2)\lambda_L$ is shown as a function of $\beta m$. This combination approaches a constant $\approx 0.025$ at small mass. Although this is no obvious from the log scale plot here, it decreases exponentially in $\beta m$ for $\beta m > 1$. We used a uniform discretization of the $u$ interval into 640 lattice points. {\bf Bottom:} the eigenvector $f_2(u)$ corresponding to $\lambda_L$ for small mass and large mass.}\label{eigenvectorPlot}
\end{center}
\end{figure}
\section{The rung function}\label{rungapp}
We will evaluate the rung function
\be
R(k) = 48\lambda^2\int \frac{d^4p}{(2\pi)^4}\,\tilde{G}(k/2+p)\tilde{G}( k/2-p).
\ee
A very similar integral was provided in \cite{jeon}. Plugging in for $\tilde{G}$ using (\ref{tildeG}) and doing the $p^0$ integral with one of the delta functions, we find
\be
\frac{3\lambda^2}{4\pi^2}\sum_{s,\tilde{s} = \pm 1}\int d^3 p \frac{\,\delta(k_0 +sE_+-\tilde{s}E_-)}{E_+E_-\sinh \frac{\beta E_+}{2}\sinh \frac{\beta E_-}{2}} \hspace{20pt} E_{\pm} = E_{\k/2 \pm \p}.
\ee
One of the four possible configurations of $s,\tilde{s}$ gives zero. The sum of the other three is 
\be\label{twoterms}
\frac{3\lambda^2}{4\pi^2}\int d^3 p \frac{\delta(|k_0| -E_+-E_-)}{E_+E_-\sinh \frac{\beta E_+}{2}\sinh \frac{\beta E_-}{2}} + \frac{3\lambda^2}{2\pi^2} \int d^3 p \frac{\delta(|k_0| +E_+-E_-)}{E_+E_-\sinh \frac{\beta E_+}{2}\sinh \frac{\beta E_-}{2}} .
\ee
At this point, we use rotation invariance to set $\k = (|\k|,0,0)$, and we decompose $\p = (p_1,p_\perp)$. The integrand depends on $p_\perp$ only through $p_\perp^2$, so we can replace $d^3p \rightarrow \pi dp_1 d(p_\perp^2)$. 

Specializing to the dispersion relation  $E_\p = \sqrt{m^2+ \p^2}$, we can write $p_1,p_\perp^2$ in terms of $E_\pm$. Scanning over all values of $p_1,p_\perp$, we cover all positive values of $E_\pm$ satisfying the constraint
\be\label{constraint}
E_+^2+E_-^2 \ge 2m^2 + \frac{\k^2}{2} + \frac{1}{2\k^2}(E_+^2-E_-^2)^2.
\ee
One also finds the Jacobian for the change of variables
\be
\frac{dp_1 d(p_\perp^2)}{E_+E_-} = \frac{2dE_+dE_-}{|\k|}.
\ee
In these variables the integral can be done straightforwardly, with the upper and lower limits of integration determined by the constraint (\ref{constraint}) and the remaining delta function. One finds that the first term in (\ref{twoterms}) is zero unless $(k^0)^2 \ge \k^2 + 4m^2$, and the second term is zero unless $\k^2\ge (k^0)^2$. All together, we find
\be\label{runganswer}
R(k) = \frac{6\lambda^2}{\pi\beta |\k|\sinh \frac{|k^0|\beta}{2}}\left[\theta(-k^2-4m^2)\log\frac{\sinh x_+}{\sinh x_-} + \theta(k^2)\log\frac{1 - e^{-2x_+}}{1 - e^{2x_-}}\right]
\ee
where $k^2 = -(k^0)^2 + \k^2$ and
\be
x_\pm = \frac{\beta}{4}\left(|k^0|\pm |\k|\sqrt{1 + \frac{4m^2}{\k^2-(k^0)^2}}\right).
\ee
Note that the variable $x_\pm$ is real whenever one of the $\theta$ functions is nonzero.

\section{Some details on finding $\lambda_L$}\label{numericsapp}

We start by converting the three-dimensional integral equation to a one-dimensional equation, for the function $f(\omega,|p|)$. In this section we will use the notation $P = |\p|$ and $K = |\k|$. To get the one-dimensional equation, we have to integrate over the angle between $\k$ and $\p$. It is simplest if we introduce a variable $y = |\k - \p|$ so that
\be
\int d^3k = 2\pi \int_0^\infty K^2dK\int_{|K-P|}^{K+P} \frac{y dy}{KP}.
\ee
The eigenvalue equation we want to solve is then
\be
\lambda_L f(P) = \int_0^\infty dK \,m_1(P,K)\left(f(K) - \frac{\sinh \frac{\beta E_P}{2}}{3\sinh\frac{\beta E_K}{2}} f(P)\right)\ee
where the angle-averaged kernel is written in terms of 
\be
m_1(P,K) = \frac{3\lambda^2}{(2\pi)^3\beta}\cdot\frac{K}{P E_P E_K}\int_{|K-P|}^{K+P}dy\left(\frac{\log\frac{\sinh x_+^+}{\sinh x_-^+}}{\sinh \frac{\beta E_+}{2}} + \frac{\log\frac{1-e^{-2x^-_+}}{1-e^{2x^-_-}}}{\sinh\frac{\beta E_-}{2}}\right).\label{angled}
\ee
In this expression, $E_\pm = |E_K \pm E_P|$ and 
\be
x^+_\pm = \frac{\beta}{4}\left(E_+\pm y\sqrt{1 + \frac{4m^2}{y^2-E_+^2}}\right) \hspace{20pt} x^-_\pm = \frac{\beta}{4}\left(E_-\pm y\sqrt{1 + \frac{4m^2}{y^2-E_-^2}}\right).
\ee
For nonzero mass, the integral in (\ref{angled}) must be done numerically. The matrix $m_1(P,K)$ is not symmetric, because of the factor of $K/P$. We will fix this below.

We now change variables from $P,K$ to $u,v$ in order to map the half-line into the unit interval, which we will then discretize uniformly. A convenient choice is
\be
P(u) = P_0\frac{u}{1-u}
\ee
where $P_0$ determines the scale of momenta that receive the most attention in the discretization. For small mass we found $P_0 \approx 3m$ to be good, and for large mass $P_0 \approx m$. Changing variables in the integral, and defining (the below contained a typo in v1)
\begin{align}
f_2(u) &= (1-u)^{-1} P(u) f(P(u))\\
m_2(u,v)&= \frac{P_0}{(1-u)(1-v)}\frac{P(u)}{K(v)}\,m_1(P(u),P(v))\\
D(u)&= \frac{1}{1-u}\frac{P(u)}{\sinh \frac{\beta E_{P(u)}}{2}}
\end{align}
the eigenvalue problem becomes
\be\label{finalint}
\lambda_L f_2(u) = \int_0^1 dv\, m_2(u,v)\left(f_2(v) - \frac{D(v)}{3D(u)}f_2(u)\right).
\ee
This is now a symmetric integral equation that can discretized as matrix multiplication. One finds a ``continuum'' of negative eigenvalues corresponding to approximately localized eigenvectors, and a discrete set of positive eigenvalues corresponding to smooth delocalized eigenvectors. $\lambda_L$ is defined as the largest positive eigenvalue. This can be determined by exact diagonalization or by iteration. For small mass, the largest negative eigenvalue is larger in magnitude than $\lambda_L$, so a simple iteration of (\ref{finalint}) will not work, but one can iterate a procedure where you update $f$ to be a weighted average of the previous $f$ and the result of applying the matrix to the previous $f$. For an appropriate choice of weighting this converges to the leading positive eigenvector. See Fig.~\ref{eigenvectorPlot} for some plots of $\lambda_L$ and the leading eigenvector.

%Unused bibitems

%\bibitem{Mueller:1994gb} 
%  A.~H.~Mueller,
%  ``Unitarity and the BFKL pomeron,''
%  Nucl.\ Phys.\ B {\bf 437}, 107 (1995)
%  doi:10.1016/0550-3213(94)00480-3
%  [hep-ph/9408245].
%  %%CITATION = doi:10.1016/0550-3213(94)00480-3;%%
%  
% 

\begin{thebibliography}{99}
\bibitem{larkin}
  A.~I.~Larkin and Y.~N.~Ovchinnikov, 
  ``Quasiclassical method in the theory of superconductivity,''
    

\bibitem{Almheiri:2013hfa} 
  A.~Almheiri, D.~Marolf, J.~Polchinski, D.~Stanford and J.~Sully,
  ``An Apologia for Firewalls,''
  JHEP {\bf 1309}, 018 (2013)
  [arXiv:1304.6483 [hep-th]].
  %%CITATION = ARXIV:1304.6483;%%
  
  

\bibitem{SS} 
  S.~H.~Shenker and D.~Stanford,
  ``Black holes and the butterfly effect,''
  JHEP {\bf 1403}, 067 (2014)
  [arXiv:1306.0622 [hep-th]].
  %%CITATION = ARXIV:1306.0622;%%
  
 

\bibitem{Shenker:2013yza} 
  S.~H.~Shenker and D.~Stanford,
  ``Multiple Shocks,''
  JHEP {\bf 1412}, 046 (2014)
  doi:10.1007/JHEP12(2014)046
  [arXiv:1312.3296 [hep-th]].
  %%CITATION = doi:10.1007/JHEP12(2014)046;%%
  
  

\bibitem{Roberts:2014isa} 
  D.~A.~Roberts, D.~Stanford and L.~Susskind,
  ``Localized shocks,''
  JHEP {\bf 1503}, 051 (2015)
  doi:10.1007/JHEP03(2015)051
  [arXiv:1409.8180 [hep-th]].
  %%CITATION = doi:10.1007/JHEP03(2015)051;%%

\bibitem{kitaev}
 A.~Kitaev, ``Hidden Correlations in the Hawking Radiation and Thermal Noise,'' talk given at at Fundamental Physics Prize Symposium, Nov. 10, 2014.
 
 Stanford SITP seminars, Nov. 11 and Dec. 18, 2014.
 
 

\bibitem{Shenker:2014cwa} 
  S.~H.~Shenker and D.~Stanford,
  ``Stringy effects in scrambling,''
  arXiv:1412.6087 [hep-th].
  %%CITATION = ARXIV:1412.6087;%%
  

\bibitem{Roberts:2014ifa} 
  D.~A.~Roberts and D.~Stanford,
  ``Two-dimensional conformal field theory and the butterfly effect,''
  arXiv:1412.5123 [hep-th].
  %%CITATION = ARXIV:1412.5123;%%
  

\bibitem{Maldacena:2015waa} 
  J.~Maldacena, S.~H.~Shenker and D.~Stanford,
  ``A bound on chaos,''
  arXiv:1503.01409 [hep-th].
  %%CITATION = ARXIV:1503.01409;%%


\bibitem{Kuraev:1977fs} 
  E.~A.~Kuraev, L.~N.~Lipatov and V.~S.~Fadin,
  ``The Pomeranchuk Singularity in Nonabelian Gauge Theories,''
  Sov.\ Phys.\ JETP {\bf 45}, 199 (1977)
  [Zh.\ Eksp.\ Teor.\ Fiz.\  {\bf 72}, 377 (1977)].

\bibitem{Balitsky:1978ic} 
  I.~I.~Balitsky and L.~N.~Lipatov,
  ``The Pomeranchuk Singularity in Quantum Chromodynamics,''
  Sov.\ J.\ Nucl.\ Phys.\  {\bf 28}, 822 (1978)
  [Yad.\ Fiz.\  {\bf 28}, 1597 (1978)].

%\cite{Fischler:2021rxy}
\bibitem{Fischler:2021rxy}
W.~Fischler, T.~Guglielmo and P.~Nguyen,
``Quantum chaos in a weakly-coupled field theory with nonlocality,''
[arXiv:2111.10895 [hep-th]].
%0 citations counted in INSPIRE as of 09 Jan 2022

\bibitem{Grozdanov:2018atb}
S.~Grozdanov, K.~Schalm and V.~Scopelliti,
``Kinetic theory for classical and quantum many-body chaos,''
Phys. Rev. E \textbf{99}, no.1, 012206 (2019)
doi:10.1103/PhysRevE.99.012206
[arXiv:1804.09182 [hep-th]].


\bibitem{jeon} 
  S.~Jeon,
  ``Hydrodynamic transport coefficients in relativistic scalar field theory,''
  Phys.\ Rev.\ D {\bf 52}, 3591 (1995)
  doi:10.1103/PhysRevD.52.3591
  [hep-ph/9409250].
  %%CITATION = doi:10.1103/PhysRevD.52.3591;%%
 

\bibitem{Forshaw:1997dc} 
  J.~R.~Forshaw and D.~A.~Ross,
  ``Quantum chromodynamics and the pomeron,''
  Cambridge Lect.\ Notes Phys.\  {\bf 9}, 1 (1997).
  %%CITATION = 00385,9,1;%%

\bibitem{kitaev2}
 A.~Kitaev, ``A simple model of quantum holography,'' talks given at KITP program ``Entanglement in Strongly-Correlated Quantum Matter,'' April 7 and May 27, 2015.

\bibitem{Sachdev:1992fk} 
  S.~Sachdev and J.~w.~Ye,
  ``Gapless spin fluid ground state in a random, quantum Heisenberg magnet,''
  Phys.\ Rev.\ Lett.\  {\bf 70}, 3339 (1993)
  doi:10.1103/PhysRevLett.70.3339
  [cond-mat/9212030].
  %%CITATION = doi:10.1103/PhysRevLett.70.3339;%%
  
  

\bibitem{parcolletgeorges}
  O.~Parcollet and A.~Georges,
  ``Non-Fermi-liquid regime of a doped Mott insulator,''
  Phys.\ Rev.\ D {\bf 59}, 5341 (1999)
  doi:10.1103/PhysRevB.59.5341.
  
  

\bibitem{Sachdev:2015efa} 
  S.~Sachdev,
  ``Bekenstein-Hawking Entropy and Strange Metals,''
  Phys.\ Rev.\ X {\bf 5}, no. 4, 041025 (2015)
  doi:10.1103/PhysRevX.5.041025
  [arXiv:1506.05111 [hep-th]].
  %%CITATION = doi:10.1103/PhysRevX.5.041025;%%


\bibitem{Parwani:1991gq} 
  R.~R.~Parwani,
  ``Resummation in a hot scalar field theory,''
  Phys.\ Rev.\ D {\bf 45}, 4695 (1992)
  [Phys.\ Rev.\ D {\bf 48}, 5965 (1993)]
  doi:10.1103/PhysRevD.45.4695, 10.1103/PhysRevD.48.5965.2
  [hep-ph/9204216].
  %%CITATION = doi:10.1103/PhysRevD.45.4695, 10.1103/PhysRevD.48.5965.2;%%
  


\bibitem{Sekino:2008he} 
  Y.~Sekino and L.~Susskind,
  ``Fast Scramblers,''
  JHEP {\bf 0810}, 065 (2008)
  doi:10.1088/1126-6708/2008/10/065
  [arXiv:0808.2096 [hep-th]].
  %%CITATION = doi:10.1088/1126-6708/2008/10/065;%%

\bibitem{contagion}
  R.~Omn{\`e}s,
``On locality, growth and transport of entanglement,'' arXiv:1212.0331.


\bibitem{husekim}
H.~Kim and D.~A.~Huse,
``Ballistic spreading of entanglement in a diffusive nonintegrable system,''
	Phys.\ Rev.\ Lett.\  {\bf 111}, no. 12, 127205 (2013).

\bibitem{Gur-Ari:2015rcq} 
  G.~Gur-Ari, M.~Hanada and S.~H.~Shenker,
  ``Chaos in Classical D0-Brane Mechanics,''
  arXiv:1512.00019 [hep-th].
  %%CITATION = ARXIV:1512.00019;%%
  
  

\bibitem{Mueller:2002gd} 
  A.~H.~Mueller and D.~T.~Son,
  ``On the Equivalence between the Boltzmann equation and classical field theory at large occupation numbers,''
  Phys.\ Lett.\ B {\bf 582}, 279 (2004)
  doi:10.1016/j.physletb.2003.12.047
  [hep-ph/0212198].
  %%CITATION = doi:10.1016/j.physletb.2003.12.047;%%

\bibitem{Jeon:2004dh} 
  S.~Jeon,
  ``The Boltzmann equation in classical and quantum field theory,''
  Phys.\ Rev.\ C {\bf 72}, 014907 (2005)
  doi:10.1103/PhysRevC.72.014907
  [hep-ph/0412121].
  %%CITATION = doi:10.1103/PhysRevC.72.014907;%%


\bibitem{Mathieu:2014aba} 
  V.~Mathieu, A.~H.~Mueller and D.~N.~Triantafyllopoulos,
  ``The Boltzmann Equation in Classical Yang-Mills Theory,''
  Eur.\ Phys.\ J.\ C {\bf 74}, 2873 (2014)
  doi:10.1140/epjc/s10052-014-2873-8
  [arXiv:1403.1184 [hep-ph]].
  %%CITATION = doi:10.1140/epjc/s10052-014-2873-8;%%
  
  



\end{thebibliography}
\end{document}